# FikoRE: 5G and Beyond RAN Emulator for Application Level Experimentation and Prototyping

Diego Gonzalez Morin, , Nokia XR Lab
Manuel J. López Morales, Universidad Carlos III de Madrid
Pablo Pérez, , Nokia XR Lab
Ana García Armada, Universidad Carlos III De Madrid
Alvaro Villegas, Nokia XR Lab

Novel and cutting-edge use cases have arisen since the first deployments of the fifth generation of telecommunication networks (5G). There are plenty of well-though optimally design 5G simulators and emulators which allow telecommunication technologies engineers and researchers to thoroughly study and test the network. However, the 5G ecosystem is not only limited to the network itself: a fast development of 5G-specific use cases can considerably accelerate the development of telecommunication technologies. We present FikoRE, our real-time Radio Access Networks (RAN) emulator carefully designed for application-level experimentation and prototyping. Its modularity and straightforward implementation allow multidisciplinary user to rapidly use or even modify it to test their own applications. In this article, we present FikoRE's architecture accompanied with relevant validation experiments and results.

## Introduction

The release of the fifth generation of telecommunication networks (5G) has produced a fast development and increased interest in novel use cases such as wireless Augmented (AR) and Virtual Reality (VR), Tele-operated Driving (ToD) or mobile Machine Learning (ML) algorithms offloading. These novel use cases rely on a nearby Multi-access Edge Computing (MEC) server to where the most computationally demanding tasks are offloaded. However, these use cases and other application's developers and researchers usually do not have access to a full 5G setup to test their solutions. This can potentially hinder the fast development of these novel technologies which can collaterally produce the development of 5G technologies to slow down. An optimal design of 5G and beyond telecommunication technologies can not happen without the input from the use cases and applications researchers and developers.

One feasible solution to accelerate these developers learning curve, which can lead to major breakthroughs, is to use Radio Access Networks (RAN) simulators. These simulators allow to study the network performance under different setups and configurations. There are plenty of RAN simulators in the state of the art, some of them widely use in research and industry. Viena Simulator[1] and NS3[2] are probably two of the most well-known simulators. They both were initially designed for LTE while they have incorporated 5G simulation capabilities over the last years. Another interesting example is Matlab 5G Toolbox [3], a very useful tool for telecommunications engineers and researchers as the flexibility provided by Matlab allows them to rapidly test their own models or algorithms. These simulators can model the network from a link level or system level point of view. The last option is particularly relevant for application layer developers and researchers, as they are not as concerned about the specific particularities modelled on the link level simulations but understand the overall system level performance of the network in different scenarios.

More interesting are the system level RAN emulators which are capable of filtering and handling actual IP traffic while accurately simulating, in real-time, a specific RAN technology or configuration. There are multiple RAN emulators in the state of the art, some of them already included as an implemented functionality in some of the simulators already mentioned, as in [2]. Besides, [4,5] also provide emulation capabilities for both 5G and LTE. From the industry side, there are other relevant emulators as the ones described in [6,7].

There is a common factor to all the mentioned emulators: they are designed to study the network behaviour. However, we could not find, to the best of our knowledge, any emulator in the state of the art specially focused on testing applications or use cases under different network setups. Most of the aforementioned emulators are extremely complex to use or modify for non-knowledgeable telecommunications engineers or researchers. These emulators can be understood as tools to study the network performance from a purely telecommunications perspective. For this same reason, they are designed with such a complexity level that considerably hinders users to modify their source code to implement and test their own solutions. Besides, these emulators learning curve can grow extremely slow for application layer developers.

There is a very interesting research attempt [8] to fulfil this gap of available emulators specifically designed for application layer researchers. While this emulator provides a simple to use RAN emulator to test different slicing and edge access approaches, its RAN implemented model is too abstract: it lacks a more accurate representation of important procedures such as the resource allocation step or channel quality information metrics estimation.

In this article we present FikoRE, a RAN emulator targeting to lower the barrier for application layer researchers to test their solutions and prototypes in real-time network simulations. The proposed emulator is capable of handling real IP traffic allowing researchers to directly test their applications on different network configurations and scenarios. FikoRE has been carefully designed to be easy to use and even modify for users with a diverse background knowledge. This is achieved by providing a high level of abstraction to allow non experts start experimenting with the tool. Despite this abstraction level, FikoRE's implementation ensures a high representability of an actual RAN deployment. Besides, the emulators architecture is extremely modular allowing straightforward source code modifications if necessary. FikoRE's end goal is to provide with a straightforward emulation tool easy to use for both experts and non-expert users.

## Emulator Structure

Our emulator is not particularly designed for understanding or testing the network, but study how the network and its different possible configurations behave for specific applications, use-cases and verticals. Consequently, we designed the emulator to comply with the following requirements:

- Work in Real-Time. Handle actual IP traffic efficiently.
- Emulate multiple users with real or simulated IP traffic.
- Simulates the real behavior of the network with sufficient accuracy.
- Besides, we wanted the emulator to be as simple to use and modify as possible which requires it to be implemented with a high modularity level allowing easy source code modifications.

As the goal is to test actual applications on different possible scenarios and understand which are the most optimal network configurations, we carefully designed the resource allocation algorithms and procedures implementation. We believe it's a crucial step in which an optimal algorithm design can produce an enhanced network performance for specific use cases or applications.

We have organized the emulator in 3 main modules:
  A. Traffic Capture/Generator (TG): this module is in charge of creating virtual traffic, according to the user-selected TG model, or capture actual IP traffic coming from a specific set of ports.
  B. MAC Layer: is in charge of the resource allocation step, implemented as an abstraction of the details described in 3GPP 38.214, 3GPP 38.211 and 3GPP 38.306 specifications.
  C. User Equipment (UE): it models a single UE, including its position estimation, according to a user-selected mobility model, channel quality estimation, and packet handling. The main modules included in each emulated UE are:
     a. Radio Link Control (RLC) and Packet Data Convergence Protocol (PDCP) Layer: this module models the behavior of the RLC and PDCP layers using a high abstraction level. Our implementation is not incorporating any particular procedure taken from the 3GPPP specifications.
     b. PHY Layer: models the main channel quality measurements, such as the Signal to Noise and Interference Ratio (SINR), Received Signal Reference Power (RSRP), Hybrid Automatic Repeat Request (HARQ) retransmissions, etc. The models implemented in this layer are described in 3GPP 38.901 specifications.

These modules are twined for each transmission direction: uplink (UL) and downlink (DL). In Fig. 1, the implemented overall data flow for each time step is depicted. We can observe the described modules and how the interact between them. The depicted workflow can be complemented with the following steps with summarize the logic behind each emulator's time step:

1. Each emulated UE's IP traffic is generated or captured by the TG module. The packets are sent to the

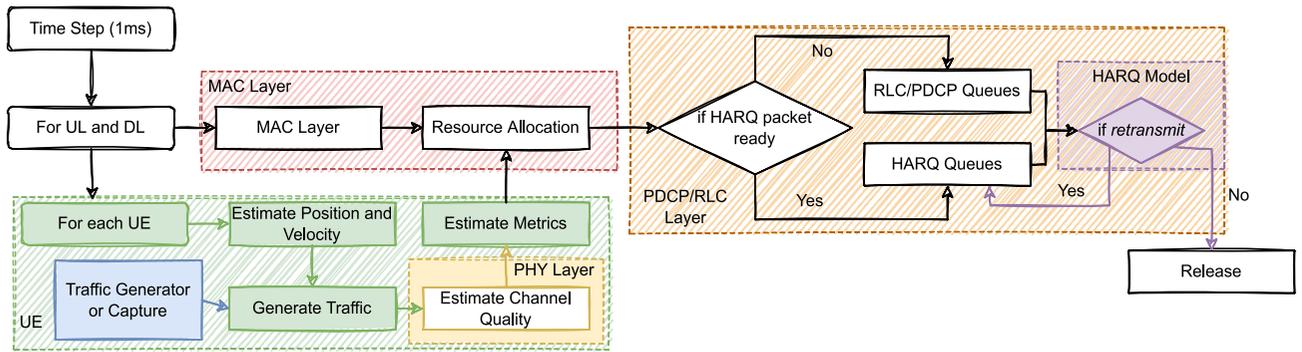

**Figure 1** *Schematic information flow between the emulator's main modules.*

RLC/PDCP layer where they are queued in the IP buffer.
2. The PDCP/RLC layer indicates the MAC layer how many bits are available for transmission for each UE via a Scheduling Request (SC).
3. With this information along with the emulated Channel Status Information (CSI) of each UE, the MAC layer allocates frequency and time resources to each UE. The CSIs are estimated by the PHY layer module.
4. The PDCP/RLC layer, according to the allocated resources, segments the IP packets into smaller blocks as determined by the MAC layer.
5. Using the emulated CSIs, the PHY layer module determines the transmission latency and HARQ retransmission probabilities of each block.
6. If the block has to be retransmitted, it is moved to the HARQ queues.
7. Finally, when the block is determined to be successfully transmitted it is moved back to the RLC/PDCP layer module, where the integrity and ordering of the full IP packets are checked.
8. Successful IP packets are then released.

The described steps run once every 1 millisecond. This deadline is strict and must be met to allow a proper synchronization between actual IP traffic and the emulator's behavior. This timing is accurately tracked by an agent running concurrently to the emulator's main threads.

The main mentioned modules are all thoroughly described in the following sections.

## Traffic Capture and Generation

The Traffic Generator (TG) module is only used for simulated UEs which are not linked with any actual IP traffic. It is in charge of periodically creating simulated IP packets for both uplink and downlink transmission directions. The current version of the emulator implements a single TG model. In this model, the user only needs to configure the target UL and DL throughputs, and the size of the simulated IP packets (in bits). Then, the TG generates, in each time step, as many packets of the specified size as necessary to fulfil the target throughput. To add some variability, we get a pseudo-random value from a normal distribution generator in each time which is added to the target DL and UL throughputs. There is a TG instance included in each simulated UE.

For UEs handling actual IP traffic, we incorporate Netfilter Queue (NQ) [9] library. NQ is a Linux user space API which provides access to packets that have been queued by the firewall. To filter packets coming from or to specific ports or addresses, the adequate firewall rules have to be set by the user. Each emulated UE linked to actual IP traffic have two unique queues assigned: one for each transmission direction (DL or UL). The appropriate queue ids are given via the configuration file to the NQ queues filtering the targeted IP packets for each UE.

Our emulator interfaces with NQ assigning a callback to each of the UE's queues, which are triggered every time a new packet has been received. The callback receives the associated IP packet id and the packet size in bits. This information, along with the callback call's time-stamp is used to create virtual IP packets identical to the ones generated by the TG module. The received packet id can be used via NQ to drop or release the associated actual IP packet.

## MAC Layer

The resource allocation step, handled by the MAC layer, determines how the frequency and time resources are divided between the connected UEs. Tuning, improving or even re-designing the main pieces in charge of the resource allocation step is necessary for the newest connected use cases and applications to become a reality. Our goal is to provide a representative yet simple to use and modify MAC layer module which allows researchers and developers to test novel scheduling schemes and implementations for their particular applications or use cases.

Consequently, we carefully designed and implemented the MAC layer module. This module implements a fairly detailed abstraction of the resource allocation procedures described in 3GPP 38.214 and 3GPP 38.211 specifications. According to such specification, the first step is to build the resource allocation grid. The resource allocation grid is divided in the frequency and time axes. The frequency axis extends along the total bandwidth and is subdivided in subcarriers which size is configurable according to the specifications. Similarly, the time axis extends along a 10 millisecond frame,

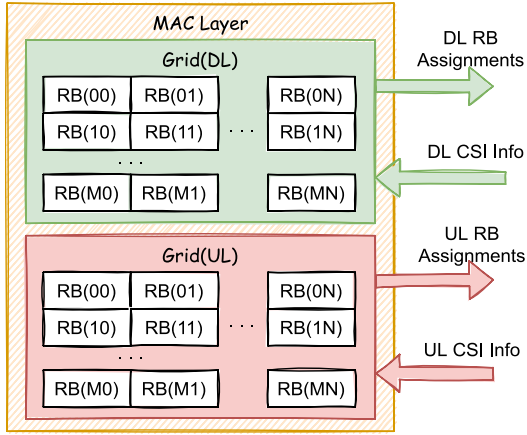

**Figure 2** *Resource allocation UL and DL grids as implemented in FikoRE emulator.*

which can be divided in sub-frames (1 milliseconds) or in even smaller units as the slots, which contain up to 14 OFDM symbols. The size in frequency of the subcarriers is determined by the numerology, which also determines the duration of an OFDM symbol and, consequently, the number of slots per sub-frame. Depending on this numerology and the size of the available bandwidth the allocation grid will have different distributions. The specific details on how the resource allocation grid is built is described in 3GPP 38.214 and 3GPP 38.211 specifications

The smallest isolated resource unit handled by the MAC layer is the resource block (RB), containing up 12 subcarriers in frequency and 14 OFDM symbols in time. Besides, the specifications also consider the aggregation of RBs in Resource Block Groups (RBGs). Our emulator implements this option, being the RBG built automatically according to the rules described in the specifications.

The main goal of the resource allocation step is to optimally assign these RBs/RBGs to the connected UEs. This step is achieved estimating, in each time-step, a metric for each UE and RB/RBG pair. Then, the PRBs/RBGs are assigned to the most optimal UEs according to a selected metric. The metric type is also selected by the user via the configuration file. The current version of the emulator implements well-known metrics such as First-In-First-Out (FIFO), Proportional Fair (PF), Blind Equal Throughput (BET) and Max. Throughput (MT).

Besides, we have implemented a simple user prioritization scheme: each UE can be configured with a different level of prioritization. This prioritization level is then used regardless the selected metric: it is applied to the estimated metric value. This feature is key in extremely demanding applications such as Virtual Reality offloading [10].

We have implemented the resource allocation model in a very modular manner so the potential users can implement their own specific resource allocation techniques, custom metrics or recent scheduling approaches such as network slicing.

The resource allocation steps is not very constraint in the specifications: the vendors and operators have a lot of design freedom, such as grouping (localized) or not (distributed) slots on the time axis.

A Resource Allocation Grid is built by the emulator for each transmission direction (UL and DL), according to the selected parameters. A simple representation of such grids is depicted in Fig. 2.

The MAC layer module is also in charge of determining which modulation to be used for each UE and PRB/RBG according to the simulated Channel Status Information (CSI) metrics estimated in the PHY layer module. This is estimated by the Adaptive Modulation and Coding (AMC) steps which dynamically adjusts the OFDM modulation order, coding method and coding rate of symbols of a specific RB/RBG to maximize throughput of the entire link transmission system. The modulation order and coding rate are directly linked with the Modulation and Coding Scheme index using the tables from the specifications. In our implementation, the MCS index is estimated and reported by the PHY layer as part of the CSI metrics. The CSI metrics can be estimated for the entire bandwidth (wideband mode) or for each sub-band, RBG or RB (sub-band mode). Both modes are implemented in our emulator. The estimated modulation order and coding rates given by the calculated MCS index is used to determine how many bits can be transmitted in each RB to the assigned UE. We used the following formula slightly modified from 3GPP 38.306:

$$bits = \nu \cdot Q_m \cdot f \cdot R \cdot N_{PRB}^{sc} \cdot B \cdot (1 - OH) \qquad (1)$$

where $\nu$ represents the number of used MIMO layers, $Q_m$ the modulation order, $f$ the scaling factor (configuration parameter), $N_{PRB}^{sc}$ the number of subcarriers in a RB, and *OH* the estimated overhead which possible values are given in 3GPP 38.306 specification. The *B* is a parameter we used to handle Time Division Duplexing (TDD), it represents how many OFDM symbols are granted for DL or UL transmission in the evaluated RB. The UL/DL OFDM symbols assignations are estimated by the TDD module, part of the MAC layer module. If Frequency Division Duplexing is selected, then *B* is always the maximum number of OFDM symbols within the RB/RBG. If RBG grouping is used, the estimated number of bits needs to be multiplied by the number of RB per RBG. Once the assignment is done, a Transport Block Size (TBS), which size is estimated according to 3GPP 38.306, is obtained for each assigned pair of UE-PRB/RBG and is moved to the correspondent UE instance. This UE instance then models all the transmission relative steps.

The key characteristic of the implemented MAC layer module in our simulator is the fact that it handles individual RB/RBG entities. Consequently, it allows to perform the resource allocation step in several manners, using different RB/RBG setups. For instance, the resource grid can be subdivided in smaller grids with different numerologies, metrics or user priorities. Our MAC layer implementation is straight-forward and extremely modular hoping that these novel techniques can be easily implemented by other potential users.

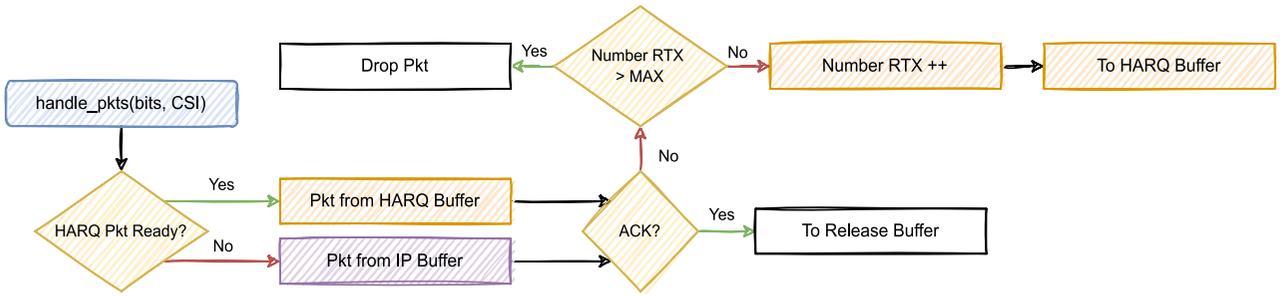

**Figure 3** RLC/PDCP Layer module's data flow simplification.

## User Equipment

The UE module models the individual UEs connected to emulated gNB. In each timestep, the UE module is implemented to:
1. Generate Traffic: using the TG model or from the incoming actual IP traffic.
2. Update Position: using our own mobility module, the UE updates its own position in every timestep. This module implements three well-known mobility models: random walk, waypoint and Manhattan.
3. Estimate the Channel State Indicators: using the PHY Layer module which implements the appropriate models from the 3GPP 38.901.
4. Handle Packets: the packets are handled by the RLC/PDCP module and can be queued, depending on several emulated metrics, in 3 different buffers: IP, HARQ and Packet Release buffers.

The main steps of the UE module are handled by the following main modules:

### User Equipment: RLC/PCDP Layer

The PDCP/RLC layer represents a simple abstraction of all the high layer levels within the full stack of 5G networks and beyond. It is in charge of queuing all the IP packets (real and simulated) from each UE, creating one IP buffer for each UE and each transmission direction (UL and DL). Using a Scheduling Request (SR), the RLC/PDCP layer communicates the MAC layer how many bits are ready for transmission for each UE. When a slot is allocated to an UE by the MAC layer, the RLC/PDCP layer splits the IP packets into smaller blocks (if necessary) of the estimated TBS. These blocks can then be moved to the HARQ or the Packet Release buffers. This is decided according to the implemented HARQ model, which models the packets retransmission probability as:

$$p(NACK) = (1 - (1 - f_{BLER}(\gamma))^{n_{tx}}) \cdot r^{n_{tx}} \quad (2)$$

where $f_{BLER}(\gamma)$ estimates the Block Error Rate (BLER) given the SINR ($\gamma$) in the time step when the packet was originally transmitted, $n^{tx}$ is the number of times the evaluated packet was already retransmitted, and $r$ is the HARQ error reduction factor which is in our emulator is left as a configuration parameter. If the packet has to be retransmitted, it is moved to the HARQ model. If, otherwise, the HARQ model determined the packet to be successfully transmitted then it is moved to the Packet Release buffer. A HARQ packet is ready to leave the buffer when the current time stamp is greater than its release time target. The release time target is estimated according to the PHY layer air transmission delays, HARQ ACK/NACK period, and other modelled processing delays. Similarly, an IP packet queued in the Packet Release buffer is released from the emulator when the release time deadline is also met. This deadline is identically estimated as in the HARQ packets. This simple approach has shown to sufficiently represent the different latencies added along the entire RAN stack, as we will analyse in the experiments section. Similarly, to the IP buffer case, there is one instance of the HARQ and Packet Release buffers for each UE.

The Packet Release buffer also implements packet re-ordering capabilities as an actual RLC layer. This is a key component as HARQ retransmission produce packet disordering. The goal of our implemented Packet Release buffer is to reorder and reconstruct the individual IP packets. If one block from an IP packet is dropped, when the maximum number of retransmissions is reached, then the entire IP packet is dropped. We achieve all these functionalities by simply ordering the queued packets by ID and checking each packet's previous packet's ID, stored at its header. This module is complex, built using multiple interconnected modules. Fig. 3 shows a simplified data flow of our RLC/PDCP layer implementation.

### User Equipment: PHY Layer

The PHY layer models the most relevant channel quality metrics estimation: SINR, RSRP, MCS and Channel Quality Indicator (CQI), among others. These metrics, besides of being continuously logged, are used by the MAC layer for performing the resource allocation step. The key metric, from which all the others are derived is the SINR. We calculate the SINR from the Received Signal Power (RSP), and the measured noise and interference. To estimate the RSP, we estimate the Pathloss and Shadowing attenuation using the models described in 3GPP 38.901 specification, given for 6 different outdoor and indoor scenarios. Besides, we also estimate and use the Rayleigh Fading. Other constant values are required but are left as configuration parameters for the user: transmission and reception antenna gains, and the transmission power.

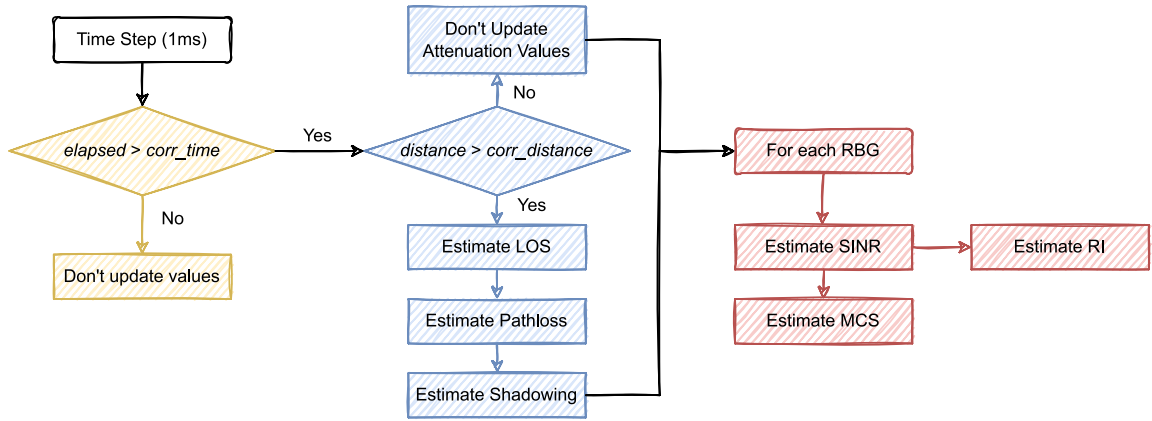

**Figure 4** *PHY Layer module's simplified data flow.*

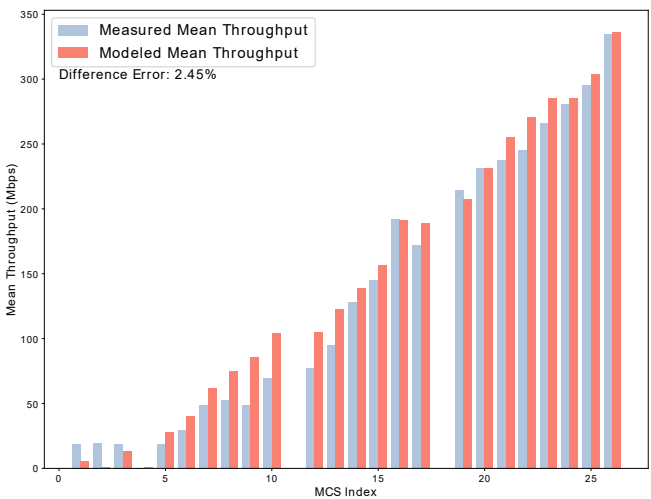

**Figure 5** *MCS index versus measured mean throughput for actual measurements on a mmW setup (blue) and simulated values(red)*

One important goal of the emulator was to allow the concurrent simulation of many users. Consequently, we needed to reduce the computation overhead as much as possible. Therefore, the noise and interference are estimated at the beginning of the emulation session and left constant. While the noise is given as a configuration parameter, the interference is estimated using the same RSP models as described before but applied to other gNBs and static UEs sufficiently close to the simulated gNBs and UEs. The number and distance of the interfering gNBs and UEs, and their base frequencies and other basic configuration are left as configuration parameters.

Once the noise, interference and RSP are estimated the SINR can be directly calculated. In the wideband mode, a single SINR value is estimated for the entire bandwidth, while in the sub-band mode, a SINR value is estimated for each bandwidth sub-band. Every how many milliseconds a new SINR value or values have to be estimated is given by the correlation time, which we estimate according to the doppler effect for the given frequency band and the UE's velocity.

The estimated SINR is used to estimate the optimal MCS index for each UE and PRB/RBG. To get the SINR-MCS relations, we performed link level simulations using Matlab, obtaining a set of SINR-BLER curves for each MCS. These are obtained for different configurations according to the maximum number of MIMO layers and RBGs sizes.

Finally, we have the Rank Indicator Model which determines if the signals received in each antenna of the UE or gNB are sufficiently uncorrelated, and MIMO can be used. As we are not estimating any correlation matrices, we proposed a simple model based in the idea that the transmitted power is constant for the same UE or gNB and is divided by each antenna used for MIMO transmission. Consequently, we can estimate the limit SINR values for each extra added MIMO layer from which is more efficient to use the extra layer. We built a simple look-up table with this information. The overall data flow of the PHY layer module is depicted in Fig. 4.

## Validation Experiments and Results

Even though the emulator has been carefully designed to be simple to use and especially easy to modify, it should provide a sufficiently accurate simulation of an actual RAN deployment. Consequently, we performed a set of thoughtful experiments to validate and show the emulator's models and performance. To ensure repeatability, all the stochastic models were substituted for static variable values. We also assumed that all the UEs were in Line of Sight (LOS) with the gNB.

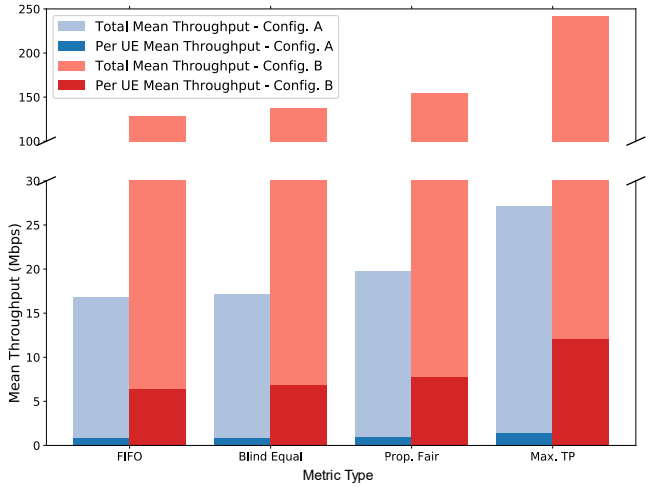

**Figure 6** *Throughput values obtained using different metric types and 20 simulated users. In red: mmW frequency band. In blue: 3.5 GHz frequency band.*

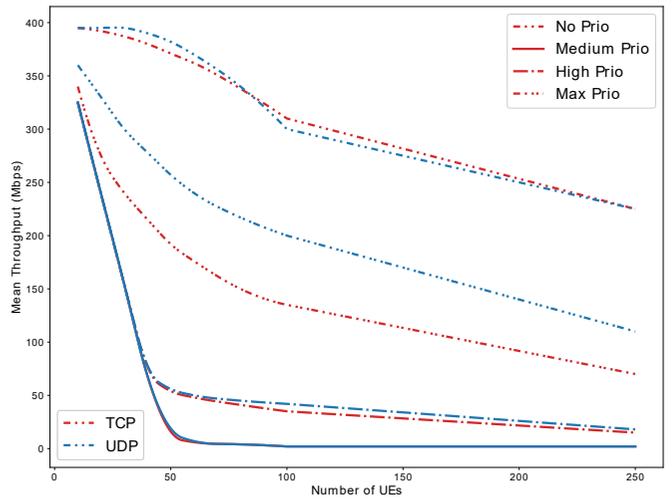

**Figure 7** *Emulated throughput for a user handling real IP data and sharing resource with other simulated UEs. Different UEs prioritization levels are used for the real IP traffic UE.*

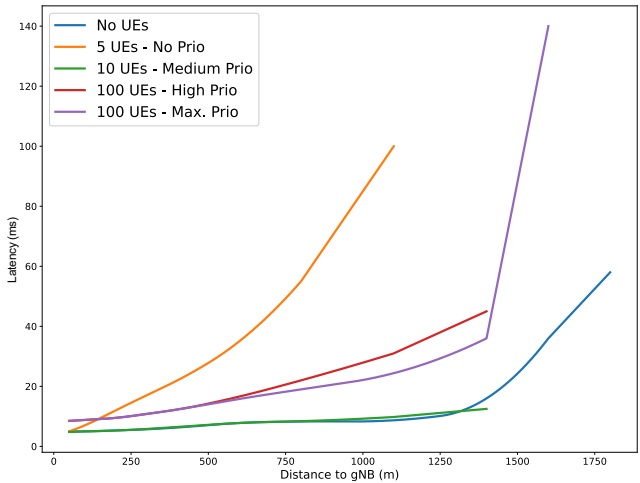

**Figure 8** *Emulated latencies for a UE mimicking a Tele-operated Driving use case. Encoded video is transmitted via RTP at a bitrate of 10 Mbps. Different number of simulated UEs, distances and user prioritizations levels are used.*

As we have already described, the MAC layer is a key component to fulfil our emulator's main goals. Therefore, the first experiment targets to validate the MCS index to allocated bits implemented model. If the relation between the MCS index and the allocated bits is off, the rest of the steps or models can not accurately represent an actual RAN setup. The main goal is to validate the used formula (1). To validate this formula, we took several measurements on an actual millilitre wave (mmW) 5G deployment. We used iperf3 to continuously transmit IP packets to an iperf3 server running on a nearby MEC. We used an Askey RTL6305 modem to wirelessly connect our laptop to the mmW gNB. We continuously logged the MAC layer level DL throughput and the associated MCS index. These pairs of MAC layer throughput and indexes were compared to the ones obtained by the emulator. Both the emulator and actual mmW setup were running on the 26.5 GHz frequency band, for a total of 800 MHz of bandwidth separated in 8 carriers. All the carriers were configured with 4DL:1UL TDD configuration. In both cases, only one UE is connected to the gNB. The results comparison is shown in Fig. 5. We can observe how in most of the cases the mean throughput difference between the actual and emulated cases its low, showing a final average error smaller than a 2.5%. We can on the contrary observed how for lower MCS indexes show more throughput differences than the mid and high indexes.

The metric estimation step of the MAC layer is also a key step as the selected or implemented metric directly affects the overall performance for particular applications. In [1] the authors tests how the overall provided throughput by the gNB is affected by the selected metric for a fixed number of connected UE. To validate our metrics and resource allocation step implementation we decided to repeat the experiment using our emulator. We tested two different setups: 3.5GHz frequency band with 40 MHz of DL bandwidth (Configuration A) and 26.5 GHz frequency band with 400 MHz of DL bandwidth (Configuration B). We tested our implementation of the FIFO, PF, BET and MT metrics. To be able to compare our results with the one obtained in [1], we configured the emulator to simulate 20 UEs demanding as much DL traffic as possible. The results obtained are depicted in Fig. 6. We can directly compare configuration A results to the one obtained in [1], showing to be almost identical. Being the Viena Simulator [1] a widely recognized RAN simulator, this comparison itself can serve as a validation of our metric estimation and resource allocation implementation.

The previous experiments were focused on the emulator's MAC layer module validation, which functionality does not distinguish between simulated or actual IP traffic. In the following experiments which evaluate the emulator's performance when actual IP traffic is filtered and used.

First, our target is to study the throughput capabilities of the emulator's handling real traffic and check how it is affected by other concurrent simulated UEs. We run the emulator configured on the 3.5 GHz frequency band and 40 MHz of DL bandwidth. The metric used is PF to ensure all the UEs, including the actual IP traffic one, are provided with a fair split of the resources. In this experiment we also target to highlight and test the user prioritization capabilities. Therefore, we repeat each experiment configuring the UE with a different prioritization level each time: no prioritization, medium, high and maximum prioritization. The other simulated UEs are in all cases set to low priority. We repeated the same experiments for both TCP and UDP traffic. The number of simulated UEs ranged from 10 to 250. The simulated UEs are demanding 5 Mbps from the network, simulating a user watching a high-definition online video. The results of the measured mean DL throughput for the actual IP traffic UE are shown in Fig. 7. As expected, the measured transmitted DL throughput rapidly decays as the number of simulated UEs grows in the medium and no prioritization cases. However, we can observe the importance of user prioritization when particular UEs much sustain a minimum network quality, such as in VR offloading [10]. When the UE is sufficiently prioritized, the measured throughput decreases at a much lower rate and sustaining high throughputs even when many UEs are demanding network resources.

Another key aspect to study is the latency simulation. We designed a simple experiment to validate this aspect of the emulator: we created a simple RTP-based video transmission pipeline which emulates a Tele-operated Driving (ToD) use case. The goal is to measure the transmission latency on the IP level, not considering the coding and encoding processing overheads. We only transmit a single video stream with a target bitrate of 10 Mbps (high quality video streaming). While more complex ToD use cases usually require multiple cameras or even a 180 or 360 camera, we choose to experiment with a single stream for simplicity. To ensure repeatability, The stream is produced using a simple high-definition video file of a first-person car driving scene. The streamed UDP packets are filtered and handled by the emulator. The UE handling the actual IP traffic stream was moving at a constant speed of 50 km/h.

From the emulator side, the setup was identical to the previous experiment. In this case, we aim to study how the latency changes depending on the distance to the gNB, the number of connected UEs, and the prioritization levels. The results obtained are depicted in Fig. 8. As expected, we can observe how the latency curves get steeper as we increase the number of simulated UEs increase and decrease the prioritization level. We can also observe that for distances further than 1300 meters, only the No UEs and Max. Prioritization cases with 100 UEs can actually transmit the stream. In all the other scenarios the IP buffer gets full, and the packets start to get lost. With this experiment we highlight the importance of user prioritization in novel use cases such as ToD.

## Conclusions

In this article, we have presented FikoER, our RAN real-time emulator specifically designed for Application-Level Experimentation and Prototyping. We have given a brief overview of other real-time emulators available in the state of the art. We have highlighted the necessity of a simple to use emulator to allow application-level researchers and engineers to test their own solutions on an emulated yet representative RAN. The described emulator aims to become an enabler of novel use cases and technologies which can not yet be tested on cutting edge technologies such as 5G's mmW frequency bands. We have thoroughly described the emulator's architecture along with the most relevant implementation details pinpointing the used models taken from the 3GPP specifications.

Finally, we have tested our emulator in several conditions in order to validate different its different modules. We have first validated the MCS index to transmitted bits formulae used by our implementation of the MAC layer. We achieve this validation by comparing actual mmW measurements with our emulate measurements. The metric estimation and resource allocation step are validated by comparing our emulated results with the results described in [1]. Finally, we have shown the emulators actual IP traffic handling capabilities in different scenarios with multiple users. In these last set of experiments, we have shown the importance of a well-designed user prioritization approach for particular demanding use cases

## References

[1] M. K. Müller et al., "Flexible multi-node simulation of cellular mobile communications: the Vienna 5G System Level Simulator," EURASIP Journal on Wireless Communications and Networking, vol. 2018, no. 1, p. 17, Sep. 2018, doi: 10.1186/s13638-018-1238-7.

[2] Riley, G.F., Henderson, T.R. (2010). The *ns-3* Network Simulator. In: Wehrle, K., Güneş, M., Gross, J. (eds) Modeling and Tools for Network Simulation. Springer, Berlin, Heidelberg. https://doi.org/10.1007/978-3-642-12331-3_2

[3] Matlab, "5G Toolbox" *5G Toolbox*, Available: https://www.mathworks.com/products/5g.html. [Accessed: 08/04/2022].

[4] Virdis, Antonio, Giovanni Stea and Giovanni Nardini. "Simulating LTE/LTE-Advanced Networks with SimuLTE." SIMULTECH (2014).

[5] G. Nardini, D. Sabella, G. Stea, P. Thakkar, A. Virdis, "Simu5G – An OMNeT++ Library for End-to-End Performance Evaluation of 5G Networks," in IEEE Access, vol. 8, pp. 181176-181191, 2020, doi: 10.1109/ACCESS.2020.3028550.

[6] Polaris Networks, " NetTest 5G Network Emulator", *NetTest 5G Network Emulator,* Available:

https://www.polarisnetworks.net/5g-network-emulators.html [Accessed: 08/04/2022].

[7] Keysight, "5G Network Emulation Solutions", *5G Network Emulation Solutions*, Available: https://www.keysight.com/zz/en/solutions/5g/5g-network-emulation-solutions.html [Accessed: 08/04/2022].

[8] J. P. Esper et al., "eXP-RAN—An Emulator for Gaining Experience With Radio Access Networks, Edge Computing, and Slicing," in IEEE Access, vol. 8, pp. 152975-152989, 2020, doi: 10.1109/ACCESS.2020.3017917.

[9] Netfilter, "libnetfilter_queue", *libnetfilter_queue,* Available: https://www.netfilter.org/projects/libnetfilter_queue/ [Accessed: 08/04/2022].

[10] D. González Morín, M. J. López-Morales, P. Perez, A., "TCP-Based Distributed Offloading Architecture for the Future of Untethered Immersive Experiences in Wireless Networks" in ACM International Conference in Interactive Media Experiences, Aveiro, 2022.

# Acknowledgements

This work has received funding from the European Union (EU) Horizon 2020 research and innovation programme under the Marie Skłodowska-Curie ETN TeamUp5G, grant agreement No. 813391.